\documentclass[%
    aip,
    jcp,
    amsmath,
    amssymb,
    twocolumn,
    superscriptaddress,
    reprint]{revtex4-1}
\usepackage{dcolumn}
\usepackage{graphicx} 
\usepackage{bm}
\usepackage{longtable}
\usepackage[english]{babel}

\begin{document}

\title{Nonequilibrium vibrational population and donor-acceptor vibrations affecting rates of radiationless transitions} 
\author{Dmitry V.\ Matyushov}
\email{dmitrym@asu.edu}
\affiliation{Department of Physics and School of Molecular Sciences, Arizona State University, PO Box 871504, Tempe, Arizona 85287}
\begin{abstract}
An analytical theory is developed for radiationless transitions in molecules characterized by nonequilibrium populations of their vibrational modes. Several changes to the standard transition-state framework follow from nonequilibrium conditions: (i) non-Arrhenius kinetics, (ii) the violation of the fluctuation-dissipation theorem (FDT), and (iii) the breakdown of the detailed balance.  The violation of the FDT is reflected in the breakdown of relations between the first (Stokes shift) and second (inhomogeneous band-width) spectral moments, and of similar relations between reorganization parameters for radiationless transitions. The detailed balance between the forward and backward rates is not maintained, requiring a lower effective free energy of the reaction relative to the thermodynamic limit. The model suggests that strong control of radiationless transitions can be achieved if a nonequilibrium population of vibrations modulating the donor-acceptor distance is produced.     	
\end{abstract}

\maketitle

\section{Introduction}
\label{sec:0}
The theory of optical transitions in molecules and crystalline impurities was established in early 1950's\cite{Huang:1950fg,Lax:52,Kubo:55,Davydov:53,Pekar:63} based on the idea that electronic energies of the light absorber couple to deformations (phonons) of the medium affecting the observed band-shape. Absorption of light by a localized impurity in the crystal creates or annihilates lattice vibrations.\cite{Huang:1950fg} An optical transition in a molecule couples to molecular normal-mode vibrations and deformation/polarization of the surrounding medium.\cite{Davydov:53,Pekar:63}  Thermal nuclear motions then drive the molecular energy levels into resonance with the radiation photon $\hbar\omega$. The coupling of the light absorbing center to the medium can be, in the leading approximation, considered as a linear function of medium displacements. This approximation has resulted in a number of closed-form solutions for optical band-shapes.\cite{Mahan:90,Mukamel:95,BixonJortner:99}   

The theory for optical transitions was later extended to radiationless transitions,\cite{Englman:1970bi,Bixon:1991gc,BixonJortner:99} which mostly follow from the original formulation in the limit of $\omega=0$. Here, one anticipates that tunneling between distinct electronic states occurs at the point of crossing of the corresponding Born-Oppenheimer (BO) surfaces. The crossing point both satisfies the energy conservation condition $\omega=0$ and the Franck-Condon principle for the tunneling of a light particle (electron or proton).\cite{Marcus:93} This general framework covers a broad range of phenomena, including electron, proton, and atom transfer reactions. The idea of crossing BO surfaces leads to an analytical theory when supplemented with linear coupling of the quantum states with thermally fluctuating nuclear modes. This formalism has enjoyed broad support by experiment and is routinely used to both calculate rates of electron/atom transfer\cite{Barbara:96} and to fit band-shapes of charge-transfer optical transitions.\cite{Chen:98}    

The importance of dynamical effects of molecular vibrations in radiationless transitions have been long recognized within the classical description of Kramers-type diffusional kinetics\cite{Sumi:86} and its extensions to quantized vibrational states.\cite{Jortner:1988jr,Barbara:1992ed} A recent revival of interest to the problem of vibronic transitions and vibrational dynamics has been driven by experiments modulating rates of electronic transitions by populating vibrational modes through infrared (IR) laser pumping.\cite{Lin:2009fr,Delor:2014aa,Petersson:2014bv,Delor:2015hm,Thomas:2016ki} Experimentally, this approach opens the door to site-selective chemistry with IR pulses.\cite{Frei:1983ez,Elles:2006kb} From the theoretical perspective, a number of molecular-scale mechanisms to affect the reaction rate can be anticipated in terms of coupling of vibrational dynamics with the electronic structure.\cite{Valianti:2018ck,Ma:2018jl} More fundamentally, nonequilibrium population of molecular vibrations is a special case of a general problem addressing activated kinetics at nonequilibrium conditions,\cite{Elles:2006kb,Craven:2016,DMpnas:16,Rafiq:2018jw} when a net energy flux through the reacting system is allowed. We show below that these conditions produce  phenomenology relevant to driven systems, including the violation of the fluctuation-dissipation theorem (FDT)\cite{Kubo:66} and the breakdown of the detailed balance.\cite{Wang:2015dj} The temperature dependence of the reaction rate, following the Arrhenius law at equilibrium conditions, becomes generally non-Arrhenius.   

The goal of this paper is to provide mathematically exact solutions for the rates of rdiationless transitions expressed through nuclear reorganization energies and effective frequencies of vibrations, which can be parametrized through more detailed calculations and interpretation of experiment. The approach adopted in this study is less geared toward specific microscopic mechanism and, instead, follows the philosophy of early studies of radiationless transitions in molecules.\cite{Huang:1950fg,Lax:52,Kubo:55,Davydov:53,Pekar:63} The focus here is on deriving closed-form solutions for Franck-Condon factors in two-state systems affected by nonequilibrium population of molecular vibrations. The present study is therefore limited to transitions described by the Fermi's golden rule (often designated as non-adiabatic reactions) and does not include solvent dynamics which  can modify the rate for each vibronic channel.\cite{Jortner:1988jr,Barbara:1992ed}       

 A closed-form expression for the rate of radiationless transitions is obtained. It generalizes the widely used Bixon-Jortner equation\cite{BixonJortner:99} through the use of nonequilibrium stationary population of vibrational states and, more significantly, by incorporating the non-Condon variation of the donor-acceptor coupling\cite{Skourtis:2010fk} induced by quantum donor-acceptor vibrations. The main result of the analytical model is to recognize that establishing a nonequilibrium population of the vibrational mode altering the donor-acceptor distance (such as bridge vibrations in donor-bridge-acceptor complexes) is the most efficient route to affect the rate of radiationless transitions through IR pumping.

\section{Theory}
\label{sec:1}
We start with considering the generic case of a vibronic transition between two parabolas shifted along a vibrational coordinate $q$. This problem was extensively studied in the past.\cite{Huang:1950fg,Lax:52,Kubo:55,Davydov:53,Pekar:63,Englman:1970bi,Bixon:1991gc,BixonJortner:99} The standard derivation of the transition probability assumes equilibrium population of vibrational modes in each electronic state. The goal of re-tracing the standard steps presented in this section is to generalize the known results to situations when populations of vibrational states are stationary, but nonequilibrium. Such a situation might occur when the molecule is exposed to sufficiently intense continuous IR radiation or to IR pulses with duration exceeding the rate of vibrational relaxation. Another window for applying this theory is for reactions faster than the rate of intermolucular vibrational relaxation to the surrounding solvent.\cite{Delor:2014aa}  

We additionally establish in this section the framework for extending the theory to the case of a vibrational normal mode modulating the donor-acceptor distance and the coupling between the initial and final states of the tunneling particle (e.g., an electron or a proton). In that latter case, the non-Condon effect of donor-acceptor vibrations modulating the donor-acceptor coupling leads to a significant modification of the standard results, which can be expressed in terms of closed-form mathematics.   

\begin{figure}
\includegraphics*[clip=true,trim= 0cm 0cm 0cm 0cm,width=5cm]{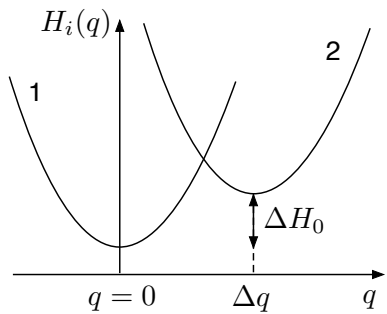}
\caption{Schematics of parabolas crossing along the effective intramolecular vibrational coordinate $q$. Parabolas represent the vibrational Hamiltonians in Eq.\ \eqref{eq1}. }
\label{fig:0}  
\end{figure}

Consider an electronic transition between two BO surfaces with minima at $q_{01}=0$ and $q_{02}=\Delta q$ along an arbitrary chosen nuclear (normal mode) coordinate $q$ (Fig.\ \ref{fig:0}). For radiationless transitions, states 1 and 2 will be identified below with, respectively, the donor and acceptor states for transfer of either the electron or the proton. Expanding the BO surfaces around the minima, one obtains in the harmonic approximation
\begin{equation}
\begin{split}
  H_1(q) &= H_{01} + \frac{k}{2} q^2,\\
  H_2(q) &= H_{02} + \frac{k}{2} (q-\Delta q)^2.\\
\end{split}
\label{eq1}  
\end{equation}
This is the standard picture of two shifted parabolas with equal force constants $k=m\omega_v^2$, $m$ is the mass. By quantizing $q$, one can write the energy gap between two surfaces as
\begin{equation}
\Delta H = \Delta H_0 +\lambda_v - \sqrt{S}\hbar\omega_v (a^\dag +a),  
\label{eq2}
\end{equation}
where $a^\dag$ and $a$ are the raising and lowering harmonic operators and 
\begin{equation}
  \lambda_v=\tfrac{1}{2} k \Delta q^2  
  \label{eq3}
\end{equation}
is the vibrational reorganization energy for the coordinate $q$. Further, $S=\lambda_v/\hbar\omega_v$ is the Huang-Rhys factor\cite{Huang:1950fg} and $\Delta H_0 = H_{02}-H_{01}$.  

The Golden-Rule probability $w(\omega)$ for the radiative transition at the photon energy $\hbar\omega$ is calculated as the first-order perturbation in the electronic/proton coupling $V$ as\cite{Lax:52,Ovchinnikov:69,Englman:1970bi,Coalson:1994gk}   
\begin{equation}
w(\omega) = \frac{2V^2}{\hbar^2}\mathrm{Re}\int_0^\infty 
dt e^{i\omega t - \frac{it}{\hbar} (\Delta H_0 +\lambda_v)+ F(t)} 
\label{eq4}
\end{equation}
with
\begin{equation}
e^{F(t)} =  \left\langle\exp\left[i\omega_v\sqrt{S}\int_0^t (a^\dag(\tau)+a(\tau)) d\tau \right] \right\rangle .
  \label{eq5}
\end{equation}
The angular brackets in this equation specify an ensemble average. Equilibrium Gibbs ensemble enters standard formulations,\cite{Lax:52,Ovchinnikov:69,Englman:1970bi} but a stationary nonequilibrium ensemble can be assumed as well. The latter choice is the meaning of the ensemble average $\langle \dots \rangle$ adopted here.  

One can further apply Bloch's identity\cite{Mahan:90} postulating that for any linear combination $c$ of raising a lowering operators $\langle e^c\rangle = e^{\langle c^2\rangle/2}$. Applying this relation to the term in angular brackets in Eq.\ \eqref{eq5} leads to the following result\cite{Lax:52,Mahan:90}
\begin{equation}
F(t) =-\omega_v^2S\int_0^t d\tau' \int_0^{\tau'} d\tau'' \phi(\tau'-\tau'')
\label{eq5-1}  
\end{equation}
with 
\begin{equation}
\phi(\tau' -\tau'') = \bar n e^{i\omega_v (\tau'-\tau'')} + (\bar n+1)e^{-i\omega_v(\tau'-\tau'')}.    
\label{eq5-2}
\end{equation}
Integration in Eq.\ \eqref{eq5-1} results in 
\begin{equation}
F(t) = i\omega_v St - S(2\bar n+1) + S\left[\bar n e^{i\omega_v t} +(\bar n+1)e^{-i\omega_vt}\right]  .
\label{eq6}
\end{equation}

In Eqs.\ \eqref{eq5-2} and \eqref{eq6} $\bar n$ is the ensemble-average population of the excited states of quantum vibrations of the mode $q$. Equilibrium ensemble does not have to be assumed, but, when this assumption is made,\cite{Lax:52,Englman:1970bi} one arrives at the textbook result\cite{Landau5} 
\begin{equation}
\bar n_\text{eq} = Q_v^{-1} \sum_{n=0}^\infty ne^{-\beta\hbar\omega_v(n+1/2)}= \left[e^{\beta\hbar\omega_v}-1\right]^{-1}.  
\label{eq7}  
\end{equation}
In this equation, $Q_v=[2 \sinh \chi_v]^{-1}$, $\chi_v=\beta \hbar\omega_v/2$ is the partition function and $\beta=(k_\text{B}T)^{-1}$ is the inverse temperature. 

Since $n=0$ does not contribute to the sum in Eq.\ \eqref{eq7}, nonequilibrium population created by an IR pulse will predominantly alter the $n=1$ term such that $\bar n=\bar n_\text{eq} +\delta n_1$. We therefore assume that $\bar n> \bar n_\text{eq}$ is created by an external source of IR radiation and will leave the population of the first excited vibrational state $n_1$ as a non-specified parameter depending on experimental conditions.    

We can next transform $\exp[F(t)]$ by using the mathematical identity\cite{Abramowitz:72}
\begin{equation}
e^{\frac{x}{2}(t+t^{-1})} = \sum_{k=-\infty}^\infty I_k(x) t^k ,  
\label{eq8}
\end{equation}
where $I_k(x)$ is the modified Bessel function. This leads to the expression\cite{Huang:1950fg,Lax:52} 
\begin{equation}
e^{F(t)} = e^{i\omega_vSt-S(2\bar n+1)}\sum_{k=-\infty}^\infty I_k[2S\sqrt{\bar n(\bar n+1)}] \left( \frac{\bar n}{\bar n + 1}\right)^{k/2} .  
\label{eq9}
\end{equation}
When substituted to Eq.\ \eqref{eq4}, one obtains for the transition rate
\begin{equation}
\begin{split}
w(\omega) =  \frac{2\pi V^2}{\hbar} &e^{-S(2\bar n+1)} \sum_{k=-\infty}^\infty I_k[2S\sqrt{\bar n(\bar n+1)}]  \\
&\left( \frac{\bar n}{\bar n + 1}\right)^{k/2}\delta\left(\hbar\omega+k\hbar\omega_v-\Delta H_0 \right) .
\label{eq10} 
\end{split}
\end{equation}

If $\bar n=\bar n_\text{eq}$, one gets $[\bar n_\text{eq}/(\bar n_\text{eq}+1)]^{(k/2)}=\exp[-k\chi_v]$ and $2S\sqrt{\bar n(\bar n+1)}=S/(\sinh \chi_v)$. At $\chi_v\gg 1$, which applies to nuclear vibrations in the quantum domain, one can use the series expansion of the Bessel function\cite{Abramowitz:72}
\begin{equation}
I_k(2S\sqrt{\bar n(\bar n+1)})\approx \frac{S^{|k|}}{|k|!}e^{-|k|\chi_v}  . 
\label{eq11}
\end{equation}
Substituting this expansion to Eq.\ \eqref{eq10}, one realizes that only terms with $k=-m$, $m>0$ substantially contribute to the sum. This observation leads to the well-established result for the probability of a radiative vibronic transition\cite{BixonJortner:99,Barbara:96}      
\begin{equation}
w(\omega) = \frac{2\pi V^2}{\hbar} e^{-S} \sum_{m=0}^\infty\frac{S^m}{m!} \delta \left( \hbar\omega- m\hbar\omega_v - \Delta H_0\right).   
\label{eq12}
\end{equation}

One can adopt a less restrictive expansion while keeping an unspecified $\bar n \ll 1$, with the result 
\begin{equation}
\begin{split}
w(\omega) = \frac{2\pi V^2}{\hbar} e^{-S(2\bar n+1)} &\sum_{m=0}^\infty\frac{[S(1+\bar n)]^m}{m!} \\
    &\delta \left( \hbar\omega- m\hbar\omega_v - \Delta H_0\right).  
    \label{eq13}
    \end{split} 
\end{equation}
Equation \eqref{eq13} does not require the equilibrium population $\bar n=\bar n_\text{eq}$ and is valid if $2S\sqrt{\bar n(\bar n+1)}\ll 1$. This condition can, however, be violated for $S>1$ and a sufficiently large $\bar n$. We will therefore keep a more general and exact result in Eq.\ \eqref{eq10} as the basis for our calculations.   

Equations \eqref{eq12} and \eqref{eq13} apply to quantum vibrations with $\beta\hbar\omega_v\gg1$. The limit of classical vibrations follows from the Taylor expansion of $F(t)$ in Eq.\ \eqref{eq6} about $t=0$, with the result $F(t)\approx - \tfrac{1}{2}\sigma_v^2(t/\hbar)^2$, $\sigma_v^2=2\bar n \lambda_v \hbar\omega_v$. Equation \eqref{eq4} is then converted to a Gaussian function
\begin{equation}
w(\omega) = \frac{V^2}{\hbar}\left(\frac{2\pi}{\sigma_v^2}\right)^{1/2}\exp\left[-\frac{(\hbar \omega - \Delta H_0 - \lambda_v)^2}{2\sigma_v^2} \right] .  
\label{eq30}
\end{equation}
When the classical equilibrium population is adopted, $\bar n = \bar n_\text{eq}=(\beta \hbar\omega_v)^{-1}$, one arrives at the standard result of the Marcus theory describing radiative and nonradiative transitions affected by  classical intramolecular vibrations.\cite{MarcusSutin}  

Equations derived so far apply to electronic transitions in a molecule in vacuum. Polar solvents add thermal noise, which is usually viewed as arising from polarization fluctuations of the thermal bath. The effect of these fluctuations on electronic states is typically described\cite{MarcusSutin} by taking the corresponding energies as linear functions of a classical Gaussian stochastic variable $X$ representing polarization fluctuations of the thermal bath.\cite{Warshel:82,Zusman:80} The transition probabilities are then obtained by adding $X$ to $\Delta H_0$: $\Delta H_0\to \Delta H_0 + X$. 

When $X$ describes thermal noise produced by collective fluctuations involving many particles of the medium, the distribution of $X$ is Gaussian (central limit theorem), with the mean $\langle X\rangle$ and the variance $\sigma_p^2=2\lambda_s k_\text{B} T$, $\lambda_s$ is the solvent reorganization energy.\cite{Marcus:1956dh,Barbara:96} By taking the average over $X$ in Eq.\ \eqref{eq10} and adopting $\omega=0$ for the radiationless transition, one obtains the nonadiabatic reaction rate (such as for electron transfer)\cite{Barbara:96}
\begin{equation}
k_\text{tr} = \frac{V^2}{\hbar} \left(\frac{\pi\beta}{\lambda_s}\right)^{1/2} \mathrm{FC} ,
\label{eq14}
\end{equation}
where
\begin{equation}
\begin{split}
\mathrm{FC} =  e^{-S(2\bar n+1)} & \sum_{k=-\infty}^\infty I_k[2S\sqrt{\bar n(\bar n+1)}]  
\left( \frac{\bar n}{\bar n + 1}\right)^{k/2}\\
&\exp\left[ -\beta \frac{(\Delta G_0 +\lambda_s + k\hbar\omega_v)^2}{4\lambda_s} \right] .
\end{split} 
\label{eq15}
\end{equation}
Here, $\Delta G_0$ is the standard free energy of the reaction. 

When the low-temperature expansion [Eq.\ \eqref{eq11}] is applied to Eq.\ \eqref{eq15}, one arrives at the well-established Bixon-Jortner equation\cite{BixonJortner:99}
\begin{equation}
  \mathrm{FC} =  e^{-S}  \sum_{m=0}^\infty \frac{S^m}{m!} 
\exp\left[ -\beta \frac{(\Delta G_0 +\lambda_s + m\hbar\omega_v)^2}{4\lambda_s} \right] . 
\label{eq16}
\end{equation}
As in Eq.\ \eqref{eq13}, this result can be corrected for an arbitrary $\bar n\ll 1$. 

In the opposite limit of classical vibrations, one starts with Eq.\ \eqref{eq30} and applies the shift $\Delta H_0\to \Delta H_0 + X$. Integration with the Gaussian variable $X$ then leads to the classical Marcus equation for the rate of radiationless transition
\begin{equation}
k_\text{tr} = \frac{V^2}{\hbar} \left(\frac{\pi\beta}{\lambda_\text{eff}}\right)^{1/2}   
\exp\left[ -\beta \frac{(\Delta G_0 +\lambda )^2}{4\lambda_\text{eff}} \right] . 
\label{eq31}  
\end{equation}

Two reorganization energies characterizing the combined effect of classical intramolecular vibrations and classical solvent fluctuations appear in Eq.\ \eqref{eq31}. The first one enters the nominator of the activation free energy
\begin{equation}
\lambda=\lambda_s+\lambda_v  . 
\label{eq32}
\end{equation}
This is the classical reorganization energy of the Marcus theory\cite{MarcusSutin,Barbara:96} combining the breadth of classical intramolecular vibrations with that of solvent polarization fluctuations into one reorganization parameter. The second reorganization energy
\begin{equation}
\lambda_\text{eff} = \lambda_s + \beta \hbar\omega_v \bar n \lambda_v
\label{eq33}   
\end{equation}
includes a generally nonequilibrium population $\bar n$. 

At equilibrium classical population $n_\text{eq}=(\beta \hbar\omega_v)^{-1}$, Eq.\ \eqref{eq33} yields the standard result, $\lambda=\lambda_\text{eff}$. On the contrary, a nonequilibrium population $\bar n>\bar n_\text{eq}$ leads to $\lambda_\text{eff}>\lambda$. This result constitutes the violation of the FDT\cite{Kubo:66} and is a special case of the  general rule that FDT is violated for electron-transfer reactions occurring under stationary but nonequilibrium conditions.\cite{DMacc:07} Gibbs ensemble does not apply in such circumstances, either because the system under study is fundamentally out of equilibrium (redox proteins\cite{DMmolLiq:18}) or because of the experimental design. It is also clear that the combination of Eqs.\ \eqref{eq31} and \eqref{eq33} leads to a non-Arrhenius dependence of the reaction rate on temperature if the dependence of $\bar n$ on temperature is distinct from $\bar n\propto T^{-1}$. Finally, the rate constants for the forward transition, $k_\text{tr}^f=k_\text{tr}$, and for the backward transition, $k_\text{tr}^b$, do not satisfy the detailed balance when $\bar n\ne \bar n_\text{eq}$. The system is out of equilibrium and there is a net flux of energy dissipating the IR pulse into the surroundings.  

By repeating the derivation leading to Eq.\ \eqref{eq31} for the backward reaction, one obtains the rate with $\Delta G_0+\lambda\to \Delta G_0 -\lambda$ in the numerator of Eq.\ \eqref{eq31}, in accordance with the standard prescriptions of the Marcus theory of electron transfer.\cite{BixonJortner:99} The values $\Delta G_0\pm\lambda$ define the average vertical transition energies. Their difference is the Stokes shift, which can be written as twice the Stokes-shift reorganization energy,\cite{DMmolLiq:18} $2\lambda^\text{St}$. The ratio of the forward and backward rates is then given in terms of the ratio $\lambda^\text{St}/\lambda_\text{eff}<1$ 
\begin{equation}
k_\text{tr}^f/k_\text{tr}^b = \exp\left[ -\beta\Delta G_0 (\lambda^\text{St}/\lambda_\text{eff})\right]  . 
\label{eq50}
\end{equation}

\begin{figure}
\includegraphics*[clip=true,trim= 0cm 0cm 0cm 0cm,width=8cm]{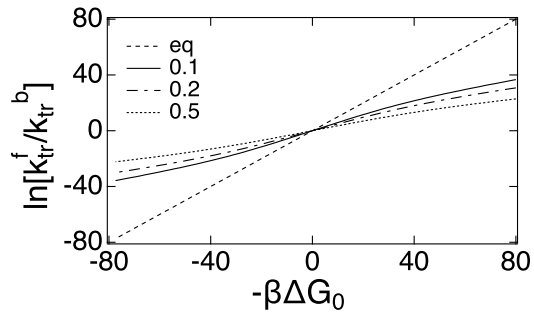}
\caption{Detailed balance condition (Eq.\ \eqref{eq50}) at varying $\bar n$. Shown is $\ln[k_\text{tr}^f/k_\text{tr}^b]$ vs $-\beta \Delta G_0$ at different values of $\bar n$ indicated in the plot. The anticipated unitary slope is obtained at $\bar n=\bar n_\text{eq}$ (dashed line). Nonequilibrium $\bar n$ values produce lower slopes in accord with Eqs.\ \eqref{eq50} and \eqref{eq51}. In contrast to those equations, obtained in the limit of classical vibrations, the calculations shown here are done for quantum intramolecular vibrations: $\lambda_s=1$ eV, $\lambda_i=0.3$ eV, and $\omega_v=2000$ cm$^{-1}$.  }
\label{fig:1}  
\end{figure}

The ratio of the rates converts to the detailed balance condition at $\lambda^\text{St}=\lambda_\text{eff}=\lambda$, when equilibrium is restored. While Eq.\ \eqref{eq50} is derived from the classical limit for the Franck-Condon factor (Eq.\ \eqref{eq31}), the same qualitative result is obtained when intramolecular vibrations are in the quantum domain and the more general Eq.\ \eqref{eq15} is used instead (Fig.\ \ref{fig:1}). Equation \eqref{eq50}, and its quantum version following from Eq.\ \eqref{eq15}, provide an experimental route for parametrizing the stationary nonequilibrium conditions achieved in the experimental design. The nonequilibrium population $\bar n$ is hard to measure directly and Eq.\ \eqref{eq50} gives access to $\bar n$ when the violation of detailed balance can be quantifies in terms of the ratio of the forward and backward reaction rates. A simple, experimentally-testable prediction of the theory is that the ratio of the forward and backward rates is below its equilibrium value for negative $\Delta G_0$ and is above it for positive $\Delta G_0$ (Fig.\ \ref{fig:1}).  

The effective reaction free energy, determined through the ratio of forward and backward rates, is reduced relative to the thermodynamic reaction free energy $\Delta G_0$ by the ratio of the Stokes and effective reorganization energies (Eq.\ \eqref{eq50}). This outcome appears to be a general result, also encountered for protein electron transfer operating at conditions of nonequilibrium sampling of configurational space by the protein.\cite{DMmolLiq:18} For system with glassy dynamics,
\begin{equation}
 \lambda^\text{St}/\lambda_\text{eff}=T/T_\text{eff} < 1  
 \label{eq51}
\end{equation}
specifies the effective temperature $T_\text{eff}$ of configurational space insufficiently sampled at the conditions of broken ergodicity.\cite{LeuzziBook,DMmolLiq:18} Below, we extend this general phenomenology to the case of radiationless transitions driven by intramolecular vibrations modulating the donor-acceptor distance. The violation of the rules of transition-state theory found here are re-enforced with stronger effects of nonequilibrium conditions on the transition rates.    

\section{Donor-acceptor vibrations and non-Condon effects}
We now take the next step in our analysis and extend it to the donor-acceptor vibrational mode as the principal (promoting) nuclear coordinate coupled to the radiationless transition (also known as Franck-Condon active mode). In this section, we assume that $q=\delta R$ describes the fluctuation of the donor-acceptor distance $R$. The new physics that this choice brings to the problem is the effect of vibrations on the donor-acceptor coupling,\cite{Valianti:2018ck} which decays exponentially with the distance
\begin{equation}
V(q) = V_0 e^{-\gamma q} .
\label{eq17}  
\end{equation}

This change of the problem requires accounting for the non-Condon effects in the calculation of the reaction rate.\cite{Skourtis:2010fk,Yang:2017ge} The rate of the radiative transition $w(\omega)$ is now written in the form
\begin{equation}
w(\omega) = \frac{2V_0^2}{\hbar^2}\mathrm{Re}\int_0^\infty 
dt e^{i\omega t - \frac{it}{\hbar} (\Delta H_0 +\lambda_v) + F(t)} ,
\label{eq18}  
\end{equation}  
where $F(t)$ changes from Eq.\ \eqref{eq5-1} to the following relation\cite{Borgis:1989hu}
\begin{equation}
F(t) = \int_0^td\tau' \int_0^{\tau'} d\tau'' f(\tau')f(\tau'') \phi(\tau'- \tau'') .
\label{eq19}  
\end{equation}
Here, $\phi(\tau'-\tau'')$ is from Eq.\ \eqref{eq5-2} and the new function $f(\tau)$ accounts for the modulation of the donor-acceptor distance by vibrations of $q$ 
\begin{equation}
f(\tau) = i\sqrt{S}\omega_v - \bar \gamma \delta(t-\tau) - \bar \gamma \delta(\tau) .
\label{eq20}  
\end{equation}
In this equation, we have introduced the dimensionless parameter of the distance decay for the donor-acceptor coupling 
\begin{equation}
\bar \gamma = \gamma |\Delta q|/\sqrt{4S} .  
\label{eq21}
\end{equation}
This parameter, in the form of an energy variable $\bar\gamma^2\hbar\omega_v$, first appeared in an analytical theory for proton and hydrogen transfer by Borgis \textit{et al.}\cite{Borgis:1989hu} However, steepest descent ansatz was applied to arrive at a closed-form solution in that work. Such approximations are avoided here, and the solution presented below is formally exact.   

Substitution of Eq.\ \eqref{eq20} to Eq.\ \eqref{eq19} leads to the following expression
\begin{equation}
\begin{split}
  F(t)=i\omega_v St &- S(2\bar n+1) -2\sqrt{S}\bar\gamma + \bar n \left(\sqrt{S}-\bar\gamma\right)^2 e^{i\omega_v t} \\
  & +(\bar n+1)\left(\sqrt{S}+\bar\gamma\right)^2e^{-i\omega_vt}  .
  \label{eq35}
  \end{split}
\end{equation}
Repeating the same steps as above for converting $F(t)$ into the series of modified Bessel functions, one can arrive at Eq.\ \eqref{eq14} for $k_\text{tr}$ with the following expression for the Franck-Condon factor
\begin{equation}
\begin{split}
&\mathrm{FC} =  e^{-S(2\bar n+1)-2\sqrt{S}\bar\gamma} \sum_{k=-\infty}^\infty I_k[2\sqrt{\bar n(\bar n+1)}|S-\bar\gamma^2|] \\ 
&\left( \sqrt{\frac{\bar n}{\bar n + 1}}\,\frac{|\sqrt{S}-\bar\gamma|}{\sqrt{S}+\bar\gamma} \right)^k 
\exp\left[ -\beta \frac{(\Delta G_0 +\lambda_s + k\hbar\omega_v)^2}{4\lambda_s} \right] . 
\end{split}
\label{eq22}  
\end{equation}
In the limit $2\sqrt{\bar n(\bar n+1)}|S-\bar\gamma^2|\ll 1$, one can again apply the series expansion for the Bessel function from Eq.\ \eqref{eq11}. If one additionally assumes $\bar n\ll 1 $, the result is a non-Condon modification of the standard Bixon-Jortner formula
\begin{equation}
\begin{split}
   \mathrm{FC} =   e^{-S-2\sqrt{S}\bar\gamma} &\sum_{m=0}^\infty \frac{(\sqrt{S}+\bar\gamma)^{2m}}{m!} \\
&\exp\left[ -\beta \frac{(\Delta G_0 +\lambda_s + m\hbar\omega_v)^2}{4\lambda_s} \right] . 
\end{split}
\label{eq23} 
\end{equation}
The non-Condon effects, caused by modulation of the donor-acceptor distance by vibrations, disappear in the limit $S\gg \bar\gamma^2$, when one returns to the Bixon-Jortner result in Eq.\ \eqref{eq16}. 
 
The limit of classical vibrations modulating the donor-acceptor distance is again obtained by the series expansion of $F(t)$ in Eq.\ \eqref{eq35} about $t=0$. The resulting expression 
\begin{equation}
k_\text{tr} = \frac{V^2}{\hbar}e^{(2\bar n+1)\bar\gamma^2} \left(\frac{\pi\beta}{\lambda_\text{eff}}\right)^{1/2}   
\exp\left[ -\beta \frac{(\Delta G_0' +\lambda )^2}{4\lambda_\text{eff}} \right] 
\label{eq38}  
\end{equation}
is similar Eq.\ \eqref{eq31} with several important changes. First, one has to replace $\Delta G_0$ with
\begin{equation}
\Delta G_0' = \Delta G_0 + 4\bar n \sqrt{\hbar \omega_v\lambda_v}\bar\gamma
\label{eq36}  
\end{equation}
and the effective reorganization energy $\lambda_\text{eff}$ in Eq.\ \eqref{eq33} with the following relation
\begin{equation}
\lambda_\text{eff} = \lambda_s + \beta (\hbar\omega_v)^2 \bar n (S+\bar\gamma^2) . 
\label{eq37}  
\end{equation}
Note that the high-temperature condition $\bar n\gg 1$ was applied in deriving both equations and $\lambda$ in the nominator in Eq.\ \eqref{eq38} is the total reorganization energy given by Eq.\ \eqref{eq32}. All statements following Eq.\ \eqref{eq31} regarding the violation of the FDT and deviations from the detailed balance apply here as well, with the new definition of $\lambda_\text{eff}$ given by Eq.\ \eqref{eq37}.  

The pre-exponential factor in Eq.\ \eqref{eq38} gains an entropic multiplier $\exp[(2\bar n+1)\bar\gamma^2]$. It arises from modulation of $V(q)$ by classical vibrations\cite{Borgis:1989hu,DMcpl:93,Kiefer:2010cp} and produces an increase in the transition rate. Due to a large value of $\gamma$ from the overlap of proton's  vibrational wave functions,\cite{Borgis:1989hu} this term in the rate preexponential factor is particularly important for proton and hydrogen atom transfer.\cite{Kiefer:2010cp} For those reactions, it accounts for the temperature dependence of the kinetic isotope effect.\cite{Hatcher:2007dd}     

\begin{figure}
\includegraphics*[clip=true,trim= 0cm 1cm 0cm 0cm,width=9cm]{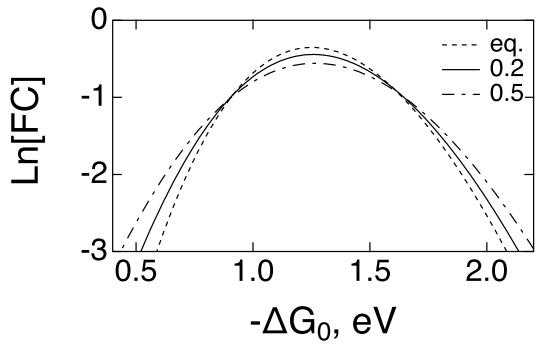}
\caption{Franck-Condon factor in Eq.\ \eqref{eq15} vs the reaction driving force $-\Delta G_0$ for reactions with varying $\bar n$: equilibrium $\bar n= \bar n_\text{eq}$ (dashed, eq.), $\bar n=0.2$ (solid), and  $\bar n=0.5$ (dash-dotted); $\lambda_s=1$ eV, $\lambda_i=0.3$ eV, and $\omega_v=1500$ cm$^{-1}$.  }
\label{fig:2}  
\end{figure}

\section{Results}
Equation \eqref{eq15} is the starting point of our analysis. It provides the rate of radiationless transition at a stationary and non-equilibrium population $\bar n$ of the effective vibrational coordinate $q$ representing the manifold of vibrational normal modes of the molecule. Given its effective character, the frequency $\omega_v$ is an average frequency\cite{DMcp:06,Yang:2014hb,Chaudhuri:2017gn} over the normal mode vibrations $\omega_i$ characterized by their corresponding reorganization energies $\lambda_i$. The equation for $\omega_v$ becomes\cite{DMcp:06}  
\begin{equation}
\omega_v = (\lambda_v)^{-1} \sum_i \omega_i \lambda_i .  
\label{eq40}
\end{equation}
Similarly, $\bar n$ is an effective average population of all vibrations populated by IR radiation and redistributed among the normal modes. The model considered here therefore assumes two time-scale separations: (i) intramolecular vibrational energy relaxation among the normal modes (with a time-scale $\tau_0$ from a few tens to hundreds of femtoseconds\cite{Arnett:99}) is much faster than the rate of the radiationless transition $k_\text{tr}$ and  (ii) the decay of $\bar n$ to $\bar n_\text{eq}$, that is intermolecular vibrational relaxation to the surrounding solvent ($\tau_{0s}\approx 10-100$ ps\cite{Elles:2006kb}), is slower than $k_\text{tr}$. Reported instances of reactions affected by IR pumping generally fall in this time window: $k_r^{-1}\approx 0.2-14$ ps\cite{Delor:2014aa} and $k_r^{-1}\approx 30$ ps.\cite{Lin:2009fr} Lifting the second approximation requires a dynamical model\cite{Feskov:2011dm,Nazarov:2016ej} for $\bar n(t)$.      

Figure \ref{fig:2} illustrates the effect of $\bar n>\bar n_\text{eq}$ on the energy gap law for radiationless transitions, which is the dependence of $k_\text{tr}$ on the driving force $-\Delta G_0$. The calculation is performed at the effective frequency of vibrations $\omega_v=1500$ cm$^{-1}$ typical for organic molecules\cite{Barbara:96,BixonJortner:99} and the vibrational reorganization energy $\lambda_v = 0.3$ eV. In addition, the solvent reorganization energy\cite{MarcusSutin} of $\lambda_s=1$ eV is  adopted. With these parameters, $\bar n_\text{eq}=7\times10^{-4}$. We show in Fig.\ \ref{fig:2} how the rate is altered when $\bar n$ exceeds $\bar n_\text{eq}$. Speeding of the reaction is seen away from the top rate of activationless transition, while the reaction becomes slower at nonequilibrium conditions near the top of the inverted parabola. Therefore, both the acceleration and slowing down of the reaction are possible depending on the driving force. This result is consistent with published experimental evidence and its theoretical interpretation.\cite{Valianti:2018ck} Creating a nonequilibrium population depletes the ground-state vibrational state. This, in turn, reduces the rate in the activationless region where the crossing of vibronic BO states involving the vibrationally ground state provides the lowest barrier. On the contrary, barrier crossing away from the activationless transition involves excited vibrational states. Increasing their population accelerates the reaction.

\begin{figure}
\includegraphics*[clip=true,trim= 0cm 0cm 0cm 0cm,width=8cm]{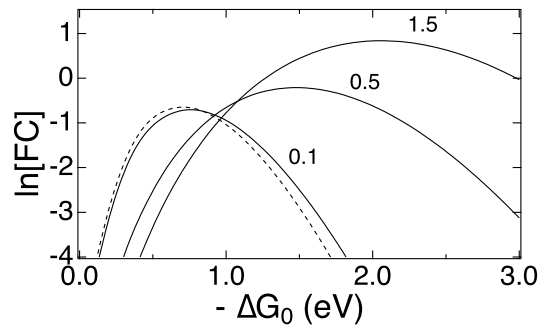}
\caption{Franck-Condon factor in Eq.\ \eqref{eq23} vs $-\Delta G_0$ with $\bar \gamma$ values indicated in the plot (solid lines). The calculations are done at $\bar n=\bar n_\text{eq}$. The dashed line indicates $\bar\gamma=0$ corresponding to the Bixon-Jortner equation [Eq.\ \eqref{eq16}]; $\lambda_s=0.5$ eV, $\lambda_v=0.3$ eV, $\omega_v=1500$ cm$^{-1}$.  }
\label{fig:3}  
\end{figure}

Turning now to vibrational modes modulating the donor-acceptor separation, the dimensionless parameter $\bar\gamma$ given by Eq.\ \eqref{eq21} is likely to be small for electron transfer. Assuming $\Delta q=0.1$ \AA, $\gamma=1.5$ \AA$^{-1}$, and $\lambda_v=0.1$ eV, one gets $\bar\gamma=0.1$ at $\omega_v=1500$ cm$^{-1}$.  This parameter is much higher\cite{Borgis:1989hu,Borgis:96,Hatcher:2007dd} for  proton and hydrogen atom transfer because of the faster distance decay of the vibrational wave functions of more massive protons yielding $\gamma = 25-35$ \AA$^{-1}$. With $\gamma=30$, our estimate yields $\bar\gamma\approx 2$ such that $\bar\gamma^2$ can exceed the Huang-Rhys factor $S$ in the power series over the vibronic transitions in Eq.\ \eqref{eq23}. Our analysis obviously applies to the regime of quantum nonadiabatic limit of proton and hydrogen atom transfer,\cite{Kiefer:2010cp} when the rate constant is calculated from Fermi's golden rule. Given a potentially broad range of $\bar\gamma$ values for different types of radiationless transitions (proton/hydrogen vs electron transfer), we show in Fig.\ \ref{fig:3} a set of curves of $\ln[\mathrm{FC}]$ vs $-\Delta G_0$ (energy-gap law, Eq.\ \eqref{eq23}) for $\bar\gamma = 0.1-1.5$ and equilibrium vibrational population, $\bar n= \bar n_\text{eq}$. 

\begin{figure}
\includegraphics*[clip=true,trim= 0cm 0cm 0cm 0cm,width=8cm]{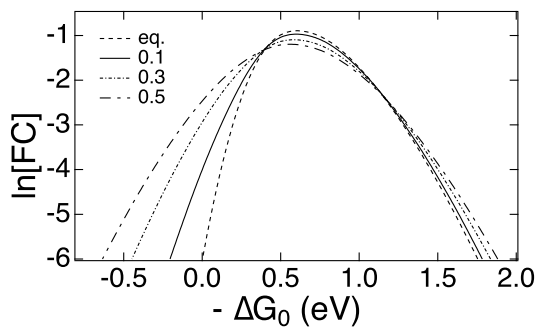}
\caption{Franck-Condon factor in Eq.\ \eqref{eq22} vs $-\Delta G_0$.  The equilibrium population is $\bar n_\text{eq}=6.8\times 10^{-5}$ (``eq'', dashed line), other lines refer to nonequilibrium populations $\bar n$ listed in the plot; $\bar\gamma=0.2$, $\lambda_s=0.5$ eV, $\lambda_v=0.3$ eV, $\omega_v=1500$ cm$^{-1}$.  }
\label{fig:4}  
\end{figure}

A significant alteration of the energy-gap law is observed with increasing $\bar\gamma$: the maximum of the distorted inverted parabola shifts to much higher driving force values and a substantial acceleration of the rate follows. These calculations suggest that $\bar\gamma>1$ makes reaching the inverted region impractical for systems typically studied experimentally. The value of $\bar\gamma$ is expected to be small for electron transfer, but, in this case, increasing $\bar n$ above the equilibrium value $\bar n_\text{eq}$ significantly accelerates the rate in the normal region $-\Delta G_0<\lambda$ (Fig.\ \ref{fig:4}).

\section{Conclusions}
This paper presents closed-form expressions for the rates of radiationless transitions in molecules characterized by stationary, nonequilibrium populations of Franck-Condon active vibrational modes. The extension of the standard Bixon-Jortner framework to nonequilibrium conditions shows a moderate change in the reaction rate. The theory's outcomes change significantly when the effective nuclear mode starts to modulate  the donor-acceptor coupling. For reactions of proton/hydrogen transfer characterized by fast distance falloff of the coupling, the non-Condon effect, even at equilibrium conditions, leads to a substantial increase in the rate in the inverted region and to a shift of the maximum of the inverted distorted parabola to higher driving force values (Fig.\ \ref{fig:3}). The non-Condon effect is much weaker for electron transfer, but here non-equilibrium conditions lead to a strong speed-up of the reaction in the normal region (Fig.\ \ref{fig:4}). The model suggests that strong control of radiationless transitions can be achieved if a nonequilibrium population of vibrations modulating the donor-acceptor distance is produced by IR pumping. 

Nonequilibrium population of molecular vibrations leads to a number of deviations from the standard transition-state theory: (i) non-Arrhenius kinetics, (ii) the violation of the FDT, and (iii) the breakdown of the detailed balance. The effective free energy of the reaction, obtained from the ratio of the forward and backward reaction rates, is reduced at nonequlibrium conditions compared to the thermodynamic limit  (Eq.\ \eqref{eq50}). For photoinduced electron transfer, part of the input energy from the radiation photon needs to be sacrificed in the form of a negative reaction free energy to achieve sufficient speedup of the forward rate and thus mostly unidirectional electron transport in natural\cite{Hoff:97} and artificial\cite{Meyer:05} photosynthesis. The model presented here suggests that ``hot'' vibrations should both accelerate photoinduced charge separation in the normal region (Fig.\ \ref{fig:4}) and reduce the effective reaction free energy dissipated by the reaction. It still remains to be seen whether this result, established from an exactly solvable model considered here, can be extended to other activated transitions at nonequilibrium conditions.

\acknowledgements
This research was supported by the Office of Basic Energy Sciences, Division of Chemical Sciences, Geosciences, and Energy Biosciences, Department of Energy (DE-SC0015641). 

\bibliography{chem_abbr,dielectric,dm,statmech,protein,liquids,solvation,dynamics,elastic,simulations,surface,nano,water,ir,et,glass,etnonlin,bioet,pt,bioenergy,photosynthNew}

\end{document}